\documentclass{webofc}
\usepackage[varg]{txfonts}   
\begin{document}
\title{Analysis of $\Xi(1620)$ resonance with chiral unitary approach}
%
%

\author{\firstname{Takuma} \lastname{Nishibuchi}\inst{1}\fnsep\thanks{\email{nishibuchi-takuma@ed.tmu.ac.jp}} \and
        \firstname{Tetsuo} \lastname{Hyodo}\inst{1}\fnsep\thanks{\email{hyodo@tmu.ac.jp}} 
}

\institute{Department of Physics, Tokyo Metropolitan University, Hachioji 192-0397, Japan
          }

\abstract{
Recently, the $\Xi(1620)$ resonance has attracted much attention, thanks to detailed experimental data  by the Belle and ALICE collaborations.
This experimental progress has prepared us to conduct theoretical analyses based on experimental data.
In this study, we analyze the $\Xi(1620)$ resonance using the chiral unitary model and discuss the properties of the $\Xi(1620)$ resonance and the $K^-\Lambda$ scattering length.
}
\maketitle
\section{Introduction}\label{intro}
The excited $\Xi$ baryons with strangeness $S=-2$ are difficult to generate experimentally, and their physical properties have not been well understood for a long time~\cite{Hyodo:2020czb}.
Recently, new detailed data are being collected, starting from the measurement of the $\pi^+\Xi^-$ invariant mass distribution in the $\Xi_c\rightarrow\pi\pi\Xi$ decays by the Belle collaboration in 2019~\cite{Belle:2018lws}, 
followed by the measurement of the correlation functions in the Pb-Pb heavy ion collisions by the ALICE collaboration in 2021, which determines the $K^-\Lambda$ scattering length~\cite{ALICE:2020wvi}.

On the other hand, in theoretical studies, the chiral unitary model is commonly used, which generates baryon resonances dynamically from the scattering of mesons and baryons~\cite{Ramos:2002xh,Garcia-Recio:2003ejq,Gamermann:2011mq,Sekihara:2015qqa,Khemchandani:2016ftn,Miyahara:2016yyh,Feijoo:2023wua,Nishibuchi:2023acl,Hai-PengLi:2023uao,Sarti:2023wlg}.
In 2002, the study in Ref.~\cite{Ramos:2002xh} has predicted the mass and width of $\Xi(1620)$, and recently, in Ref.~\cite{Feijoo:2023wua}, the analysis with the higher order terms in the chiral Lagrangian has also  been performed.
However, the width of $\Xi(1620)$ obtained in these studies is broader than the result of Belle, and the $K^-\Lambda$ scattering length determined by ALICE is not used to constrain the theoretical models.
In this study, we construct the model with a narrow decay width of $\Xi(1620)$ which is implied by the result of Belle. We also construct a model which reproduces the scattering length determined by ALICE to investigate the nature of $\Xi(1620)$. Details of this work can be found in Ref.~\cite{Nishibuchi:2023acl}.

\section{Formulation}\label{sec:Form}
The coupled-channel meson-baryon scattering amplitude $T_{ij}(W)$ with the total energy $W$ is given by the interaction kernel $V_{ij}(W)$ and the loop function $G_i(W,a)$, which satisfy the following scattering equation
\begin{align}\label{eq:teq}
T_{ij}(W)=V_{ij}(W)+V_{ik}(W)G_{k}(W,a)T_{kj}(W).
\end{align}
Indicies $i,j$ denote the meson-baryon channel.
We adopt the Weinberg-Tomozawa term for $V_{ij}(W)$, which is an S-wave interaction satisfying the chiral low-energy theorem, and we use $G_i(W,a_i)$ with the dimensional regularization to remove the divergence of the loop function.

The $\Xi(1620)$ resonance is coupled to four channels, $\pi\Xi$, $\bar{K}\Lambda$, $\bar{K}\Sigma$ and $\eta\Xi$ in the isospin basis.
Because the Weinberg-Tomozawa term $V_{ij}(W)$ with no free parameter is determined  only by chiral symmetry, the free parameters in this model are the subtraction constants $a_i$, which correspond to the cutoff parameter of the loop momentum.
In the calculation of this study, there are four coupled channels, and we construct models by choosing four subtraction constants.

\section{Numerical result}\label{sec:Num}
\subsection{Model 1}\label{subsec:model1}
In the previous study of $\Xi(1620)$~\cite{Ramos:2002xh},  $a_i$ in all channels are set to be $-2$ to match the standard cutoff size, resulting in the pole at $W=1607-140i\ {\rm{MeV}}$, identified as $\Xi(1620)$.
On the other hand, Belle reported the mass and width of $\Xi(1620)$ as $M_R=1610$ MeV and $\Gamma_R=60$ MeV~\cite{Belle:2018lws}.
In this study, we assume that the pole of $\Xi(1620)$ locates at
\begin{align}\label{eq:polest}
z=1610-30i\ {\rm{MeV}}\ ,
\end{align}
based on the Belle result. This pole appears below the $\bar{K}\Lambda$ threshold and identified as a quasibound state~\cite{Nishibuchi:2023acl}. We search for the model which reproduces the assumed pole position.
Following Ref.~\cite{Ramos:2002xh}, we set $a_{\bar{K}\Sigma}=a_{\eta\Xi}=-2$.
We bring the pole closer to the assumed one~\eqref{eq:polest} by adjusting the subtraction constants in the $\pi\Xi$ and $\bar{K}\Lambda$ channels.
As a result, at $a_{\pi\Xi}=-4.19$ and $a_{\bar{K}\Lambda}=-0.14$, the assumed pole position is reproduced with an accuracy of 1\ MeV, and a model with the quasibound state suggested by the Belle result is constructed.
Hereafter, this model is referred to as Model 1.

\begin{figure}[tbp]
\centering
\includegraphics[width=6cm]{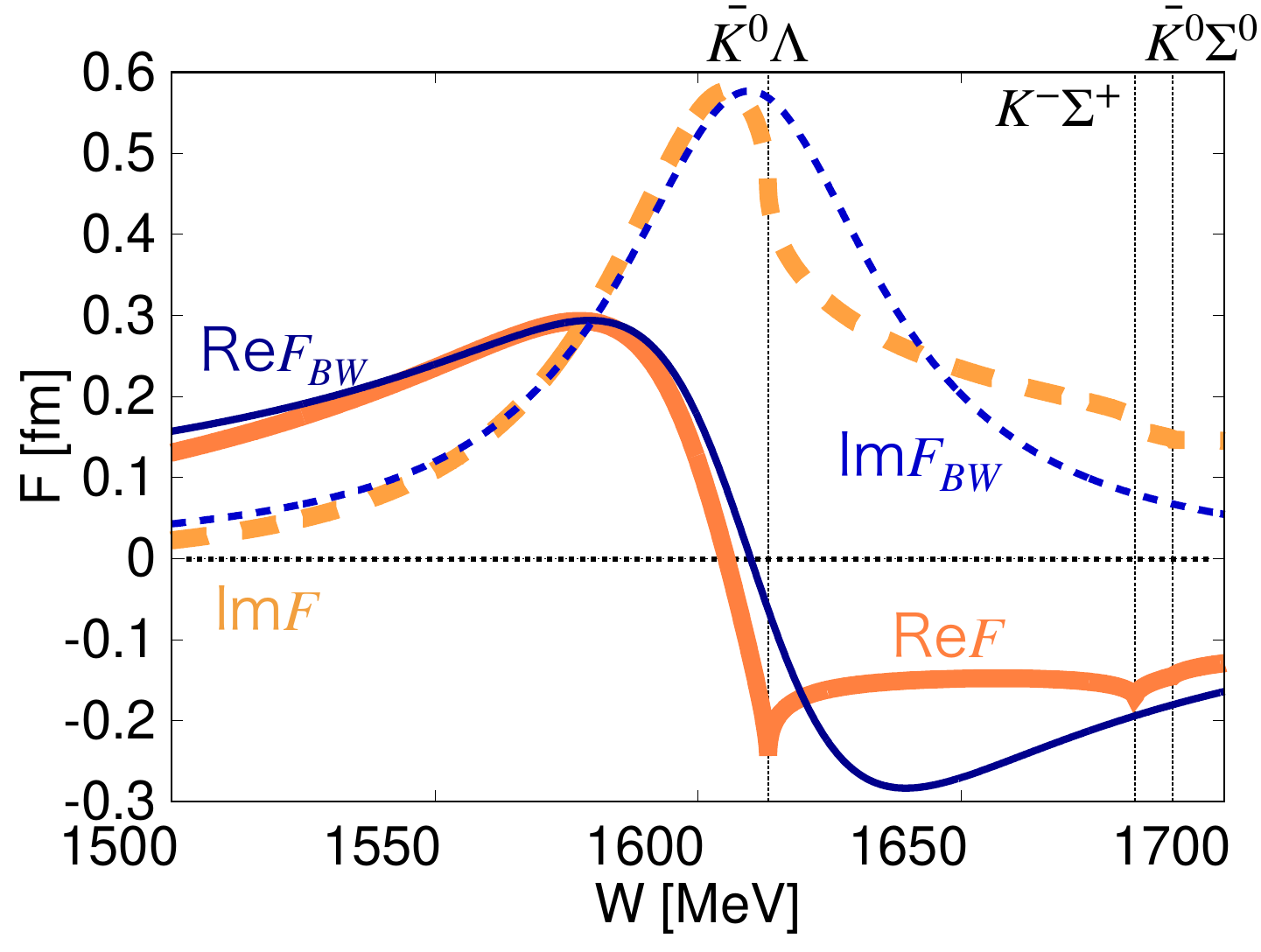}
\includegraphics[width=6cm]{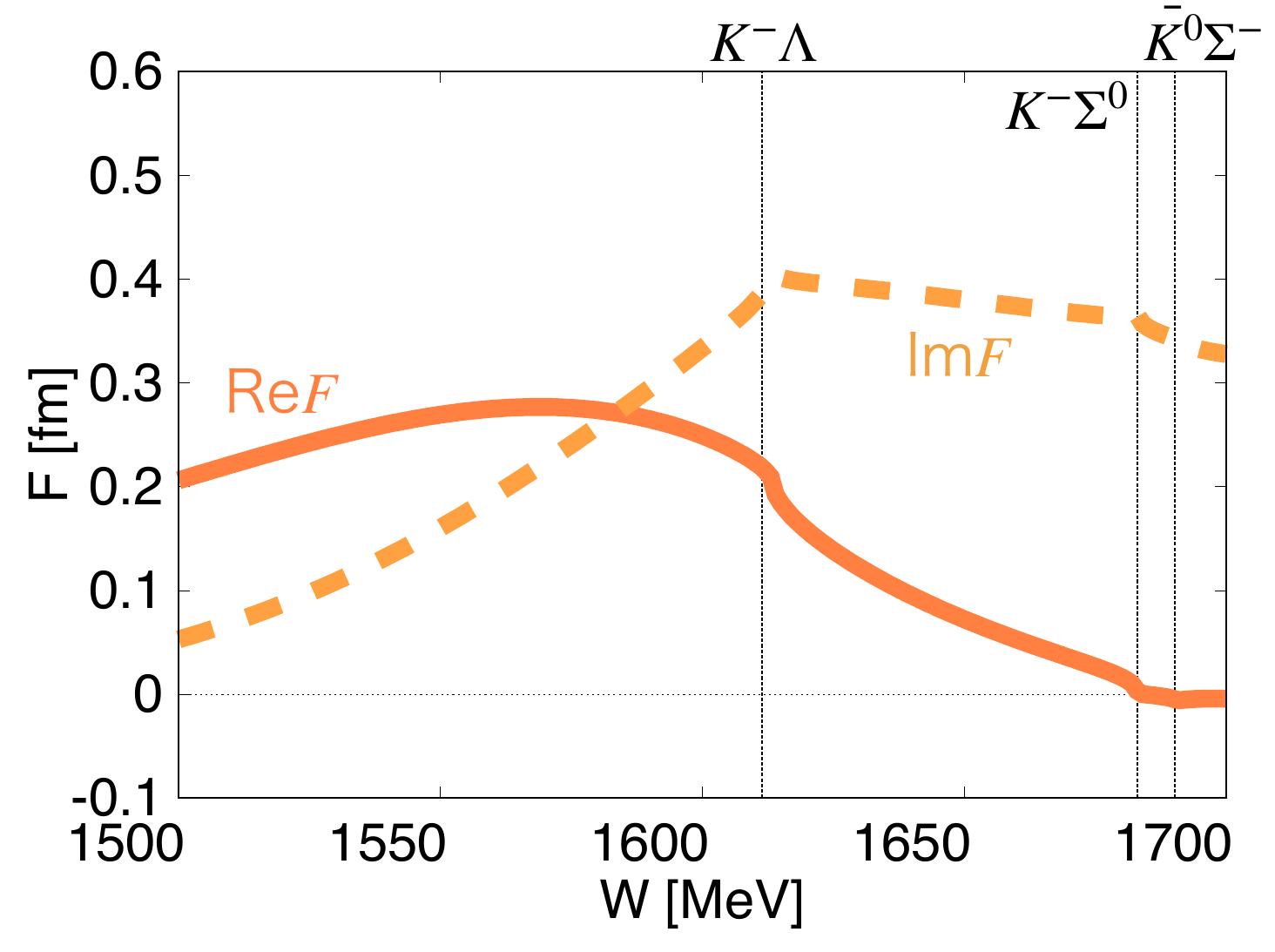}
\caption{Real part (solid line) and imaginary part (dashed line) of the meson-baryon elastic scattering amplitudes as the functions of the energy $W$.
Left : Comparison of the $\pi^+\Xi^-$ scattering amplitude of Model 1 (thick lines) with the Breit-Wigner amplitude (thin lines).
Right : the $K^-\Lambda$ scattering amplitude of Model 2.
}
\label{fig-1}       
\end{figure}

In the left panel of Fig.~\ref{fig-1}, we plot the elastic $\pi^+\Xi^-$ scattering amplitude of Model 1 together with the Breit-Wigner amplitude which has the pole at the same position.
We find that a distinct peak of the imaginary part of Model 1 appears, as in the invariant mass distribution of the Belle result.
On the other hand, comparing with the Breit-Wigner amplitude, we find that the amplitude is distorted and the peak position is shifted near the threshold of $\bar{K}\Lambda$.
Thus, the threshold effect should be taken into account for the quasibound states near the threshold.

\subsection{Model 2}\label{subsec:model2}
In the ALICE experiment, the central value of the $K^-\Lambda$ scattering length $a_0$ has been determind as~\cite{ALICE:2020wvi}
\begin{align}\label{eq:sclength}
a_0=-0.27-0.40i\ {\rm{fm}}.
\end{align}
Because the $\Xi(1620)$ resonance is located near the $K^-\Lambda$ threshold at $1609.4$ MeV, the $K^-\Lambda$ scattering length can strongly constrain the scattering amplitude near $\Xi(1620)$.

As in Model 1, we set $a_{\bar{K}\Sigma}=a_{\eta\Xi}=-2$, and $a_{\pi\Xi}$ and $a_{\bar{K}\Lambda}$ are adjusted to reproduce the $K^-\Lambda$ scattering length obtained in the ALICE experiment.
The optimization of the scattering length results in $a_{\pi\Xi}=-2.90$ and $a_{\bar{K}\Lambda}=0.36$ which reproduce the $K^-\Lambda$ scattering length~\eqref{eq:sclength} in the accuracy of $0.01$ fm. We call this Model 2.

We show the $K^-\Lambda$ scattering amplitude of Model 2 in the right panel of Fig.~\ref{fig-1}.
The imaginary part of the scattering amplitude has a cusp at the threshold energy of $K^-\Lambda$,  without showing a clear peak. This result is qualitatively different from the amplitude of Model 1 (left).
We also search for the pole in the complex energy plane, but finding no pole on the physically relevant Riemann sheet.

To find poles on the other Riemann sheets, we estimate the pole position using the scattering length $a_0$.
Based on the effective range expansion, the pole position $z$ is estimated to be
\begin{align}
z\sim\frac{-1}{2\mu_{K^-\Lambda}a_0^2}+M_\Lambda+m_{K^-}, \label{eq:zsim}
\end{align}
if $|a_0|$ is sufficiently large.
Substituting the scattering length determined by the ALICE experiment into Eq.~\eqref{eq:zsim}, the pole position is estimated to be $1701+228i$ MeV on the [ttbttt] sheet, where t and b represent first and second Riemann sheet.
In fact, we find that Model 2 has a pole in the [ttbttt] sheet. 
A pole in this Reimann sheet is called the quasivirtual state~\cite{Nishibuchi:2023acl}.

\subsection{Comparison}\label{subsec:comp}
\begin{figure}[tbp]
\centering
\includegraphics[width=6cm]{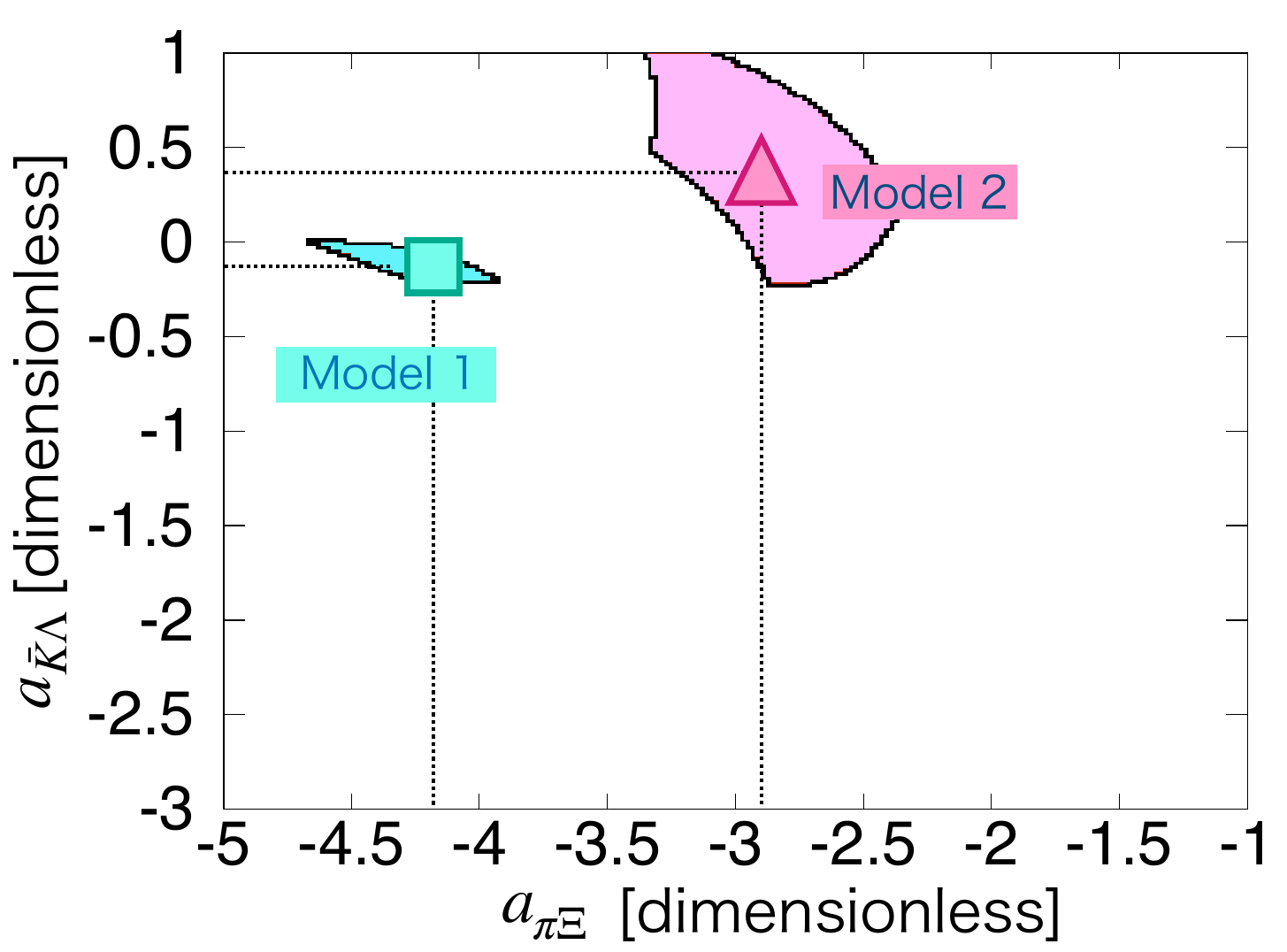}
\caption{The regions of subtraction constants of $a_{\pi\Xi}$ and $a_{\bar{K}\Lambda}$  for Model 1 and Model 2 with errors taken into account.
}
\label{fig-2}       
\end{figure}

Both Model 1 and Model 2 are based on the experimental data, but the obtained subtraction constants $a_{\pi\Xi}$ and $a_{\bar{K}\Lambda}$ are different.
In this section, we search for models with subthreshold pole suggested by the Belle result,
while reproducing the scattering lengths of the ALICE experiment by also accounting for the experimental errors.
We consider the sum of squares of statistical and systematic errors for both experiments.
In Fig.~\ref{fig-2}, we show the regions of subtraction constants $a_{\pi\Xi}$ and $a_{\bar{K}\Lambda}$  for Model 1 and Model 2, taking into account the errors.
From this figure, we find that there is no set of subtraction constants that satisfies both the constrains, because two regions have no common parts.
However, the pole position in Model 1 is assumed from the Breit-Wigner fit to the $\pi^+\Xi^-$ invariant mass distribution.
Because of this assumption, it is not necessarily concluded that the Belle result is not compatible with ALICE.  In analyzing the results of the Belle result, it is appropriate to compare directly the model with the $\pi^+\Xi^-$ invariant mass distribution without assuming the pole position.

\section{Summary}
\label{sec:Sum}
In this study, we have performed a theoretical analysis of $\Xi(1620)$, using the chiral unitary model based on the results from the Belle and ALICE experiments.
In section \ref{subsec:model1}, by adjusting the subtraction constants, we have constructed Model 1 which reproduces the assumed pole position from the Belle result. We find the threshold effect for the $\Xi(1620)$ peak below the $\bar{K}\Lambda$ threshold. 
In section \ref{subsec:model2}, we construct Model 2 which reproduces the $K^-\Lambda$ scattering length determined by the ALICE experiment. We show that there is no pole of $\Xi(1620)$ in the physically relevant Riemann sheet. Instead, a quasivirtual pole is found at the [ttbttt] sheet.
In section \ref{subsec:comp}, we show that Model 1 and Model 2 are incompatible, even with accounting for the experimental errors.
This result suggests that $\Xi(1620)$ as a shallow quasibound state below the threshold is incompatible with the $K^-\Lambda$ scattering length from the ALICE experiment.

\end{document}